\documentclass[aps,prl,twocolumn,showpacs,amssymb]{revtex4}
\usepackage{graphicx}
\usepackage{hyperref}
\usepackage{natbib}
\usepackage{bm}
\begin{document}
\title{Decoherence by a chaotic many-spin bath}
\author{J. Lages}
\affiliation{Ames Laboratory and Department of Physics and Astronomy, 
  Iowa State University, Ames IA 50011, USA}
\author{V. V. Dobrovitski}
\affiliation{Ames Laboratory and Department of Physics and Astronomy, 
  Iowa State University, Ames IA 50011, USA}
\author{B. N. Harmon}
\affiliation{Ames Laboratory and Department of Physics and Astronomy, 
  Iowa State University, Ames IA 50011, USA}
\begin{abstract}
We numerically investigate decoherence of a two-spin system (central
system) by a bath of many spins 1/2. By carefully adjusting
parameters, the dynamical regime of the bath has been varied from
quantum chaos to regular, while all other dynamical
characteristics have been kept practically intact. 
We explicitly demonstrate that
for a many-body quantum bath, the onset of quantum chaos leads to
significantly faster and stronger decoherence compared to an equivalent
non-chaotic bath. Moreover, 
the non-diagonal elements of the system's density matrix
decay differently for chaotic and non-chaotic baths. Therefore,
knowledge of the basic parameters of the bath (strength of the 
system-bath interaction, bath's spectral density of states)
is not always sufficient, and 
much finer details of the bath's dynamics can strongly affect the
decoherence process.
\end{abstract}

\maketitle

Real physical systems are never isolated.
Interaction of a quantum system with its environment 
leads to decoherence:
the initial pure state of the system quickly decays into an
incoherent mixture of several states \cite{vonNeumann,gardiner}.
Modern experiments provide much information about the 
decoherence dynamics of single (or few)
ions \cite{singleion}, Cooper pairs \cite{nakamura}, 
or spins \cite{mrfmraffi},
and require a comprehensive theory for adequate understanding.
Decoherence is also a major obstacle to building
a practical quantum computer, which, for a wide class of problems,
is exponentially more efficient than classical computers \cite{qucomp}.
Interaction of a quantum computer with the bath leads to a fast generation of errors, 
and an accurate theory is needed to find a way of controlling this process.

Decoherence is a complex quantum many-body phenomenon,
and its detailed description is a challenging problem. 
Many theoretical approaches 
eliminate the environment from consideration, approximating its
influence by suitably chosen operators (deterministic
or stochastic), and retaining only basic information:
the strength of the system-bath interaction,
characteristic energies/times of the bath, etc. \cite{gardiner} 
Such methods often work well, but 
many situations require detailed account of the bath's internal
dynamics. Recently, the role of quantum chaos \cite{quchaos} in the decoherence
process has become a subject of debate 
%\cite{zurekchaos,Alicki,JacquodAdagBeen,Srednicki,LEgen}. 
\cite{zurekchaos,Alicki,LEgen}. 
Qualitative semiclassical arguments indicate
that the chaotic bath (i.e., the bath having only a few
trivial integrals of motion) is ``a stronger decoherer'' \cite{zurekchaos}
than an integrable bath (i.e., the bath possessing a complete set
of the integrals of motion). Perturbative arguments lead to the
opposite answer \cite{Alicki}.
However, these and related works \cite{LEgen},
eliminate the central system from discussion, considering instead 
a static perturbation acting on a bath, and/or treat the bath
semiclassically, 
as a particle (or a single large spin) with an integrable or a chaotic Hamiltonian.

Although many valuable insights have been obtained in previous work,
an important question remains unanswered: is the onset of 
quantum chaos important for the real-world
situation when both the system and the bath are fundamentally
quantum many-body objects with non-trivial dynamics?
In this paper, we give an affirmative answer to this question. 
In contrast with previous work, we do not replace the system or the
bath by perturbation. We go beyond the semiclassical one-body
description, realistically considering the spin environment as many
interacting spins 1/2, which have no well-defined semiclassical limit.
We show that the chaotic bath decoheres the central system stronger
and faster than an equivalent non-chaotic one, and changes the dynamics of the
decay of non-diagonal elements of the system's density matrix.

\begin{figure}
\includegraphics[width=8cm]{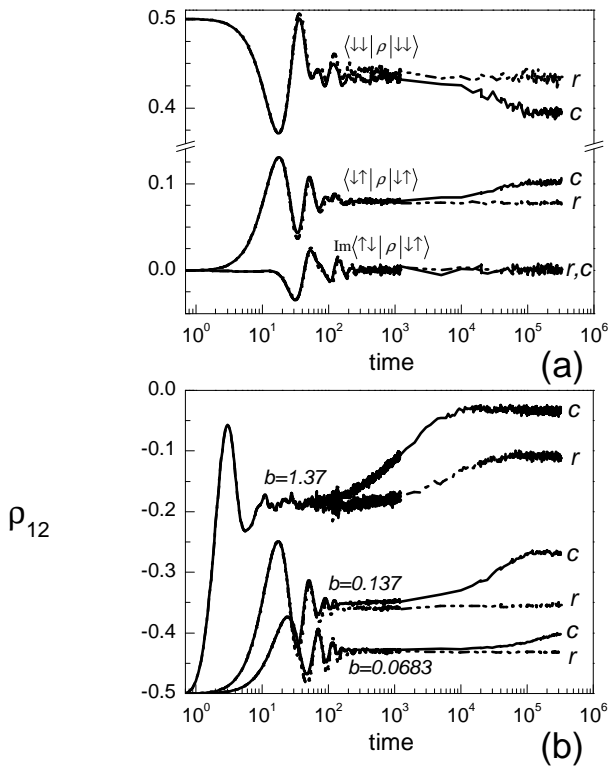}
\caption{\label{fig1} Time evolution of the elements
of the system's density matrix $\rho$ for a chaotic $\Gamma=0.04$, $h=0.014$
(\textsf{\emph{c}}) and a regular $\Gamma=0.008$, $h=0.014$
(\textsf{\emph{r}}) environment. (a) The dynamics of the elements
$\langle\downarrow\downarrow|\rho|\downarrow\downarrow\rangle$, 
$\langle\downarrow\uparrow|\rho|\downarrow\uparrow\rangle$, and
${\rm Im}\langle\uparrow\downarrow|\rho|\downarrow\uparrow\rangle$
for $b=0.137$. (b) The evolution of the element
$\rho_{12}={\rm Re}\langle\uparrow\downarrow|\rho|\downarrow\uparrow\rangle$
From top to bottom the coupling
between central system and environment is $b=1.37, 0.137, 0.0683$.
The coupling energy between the two central spins is $J=0.1$.
Everywhere below, the energy and time quantities are dimensionless.}
\end{figure}

The bath of spins 1/2 (nuclear or electron spins, magnetic impurities) 
constitutes a major source of decoherence for 
nuclear magnetic resonance (NMR) experiments, 
decoherence of phosphorus spins in Si \cite{dassarma}, 
spins in magnetic molecules \cite{stamprok} and
quantum dots \cite{loss}. Two-level defects, 
governing decoherence in Josephson junctions \cite{jj}, can 
also be modeled as spins 1/2. Even small coupling between
the bath spins can make the bath chaotic, and we need
to understand, at least qualitatively, how this affects the decoherence process.
The dynamics of a system decohered by the spin bath is
affected by many factors.
In order to conclusively separate the impact of chaos in the bath,
and to provide the knowledge needed for more complex studies,
we need a simple, well-characterized, but realistic model.
Here, we consider a central system of two exchange-coupled
spins 1/2, ${\bf S}_1$ and ${\bf S}_2$, where ${\bf S}_1$ 
interacts with a bath of spins ${\bf I}_k$ ($I_k=1/2$, $k=1,\dots N$). 
The corresponding Hamiltonian is
\begin{equation}
H = J{\bf S}_1{\bf S}_2 + {\bf S}_1\sum_k A_k{\bf I}_k + H_B
\label{sys}
\end{equation}
where $A_k$ are the system-bath coupling constants, and $H_B$
is the Hamiltonian of the bath. 
Similar models  
describe cross-relaxation and double resonance in NMR \cite{abragam},
and destruction of the Kondo effect by decoherence \cite{ourkondo}.
Detailed theoretical assessment of 
specific experiments requires separate consideration,
beyond the scope of this paper, but this simplified
model captures essential physical details of decoherence by
the chaotic spin bath. 
Due to similar reasons, we describe the bath by the 
``spin glass shard'' model \cite{georgeot} with the Hamiltonian
\begin{equation}
H_B = \sum_{k,l} \Gamma_{kl} I^x_k I^x_l + \sum_k h^z_k I^z_k +
  \sum_k h^x_k I^x_k
\label{bath}
\end{equation}
with random $\Gamma_{kl}$ and $h^{x,z}_k$, uniformly distributed
in the intervals $[-\Gamma_0,\Gamma_0]$ and $[0,h_0]$
respectively. This model 
describes the regular-to-chaotic transition in a
simple and clear way, and permits straightforward control of the
bath's dynamics \cite{georgeot}. For small $\Gamma_0$, the
bath is integrable, and becomes chaotic for 
$\Gamma_0 > \Gamma_{cr}\sim h_0/(zN)$ where $N$ is the
number of bath spins and $z$ is the number of neighbors coupled
via the term $\Gamma_{kl} I^x_k I^x_l$. Therefore, in real
baths with large $N$, the chaotic regime can be relevant even for very
small couplings $\Gamma_{kl}$. 

We study decoherence by numerically solving the time-dependent Schr\"odinger
equation for the wave function $|\Psi(t)\rangle$ of the compound system 
(the central system plus bath),
using the Hamiltonians (\ref{sys})--(\ref{bath}), and
considering up to $N=16$ bath spins (the results do not change much
already for $N>10$).
We use Chebyshev's polynomial expansion, 
in order to work with large Hilbert spaces and to
study the system's dynamics at extremely long times \cite{mehans,hansme}.
The initial state of the compound system 
is $|\Psi(0)\rangle = |\phi\rangle|\chi\rangle$, where the
state of the central system is maximally entangled,
$|\phi\rangle=\frac{1}{\sqrt{2}}(|\!\uparrow\downarrow\rangle-|\!\downarrow\uparrow\rangle)$ 
(singlet),
and the state of the environment $|\chi\rangle$ 
is a superposition of all basis states with random coefficients
(which corresponds e.g.\ to the bath of nuclear spins at temperatures
above few tenths of Kelvin).
A wide range of the parameters
$J$, $h_0$, $\Gamma_0$, $N$, and different sets of $A_k$ have been explored,
and typical results are presented below.

\begin{figure}
\includegraphics[width=8cm]{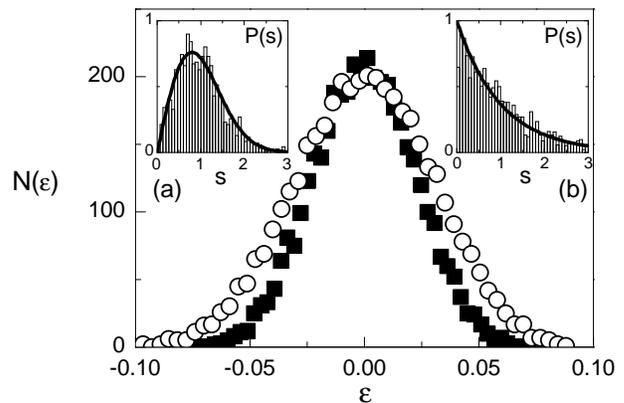}
\caption{\label{fig2} The spectral densities of states $N(\epsilon)$
vs.\ energy $\epsilon$ for regular bath ($\blacksquare$) with
$\Gamma_0=0.008$, $h_0=0.014$, and for chaotic bath ($\circ$) with
$\Gamma_0=0.04$, $h_0=0.014$. 
Insets (a) and (b) show the level spacing
distributions $P(s)$ for chaotic and regular baths, respectively.
The thick lines show the Wigner-Dyson and
Poisson distributions for chaotic and non-chaotic baths, respectively.}
\end{figure}

It is convenient to describe the system's evolution by the reduced
density matrix $\rho(t)={\rm Tr}_B |\Psi(t)\rangle\langle\Psi(t)|$
where ${\rm Tr}_B$ means trace over the bath states.
Dynamics of some elements of $\rho(t)$ is shown in Fig.~\ref{fig1}a.
Two stages are clearly seen: first,
the bath rapidly decoheres the system, excites the triplet states,
and the system oscillates between the singlet and triplet states.
Much later, thermalization takes place at much slower rate (note
the log scale of the time axis). 
Fig.~\ref{fig1}b shows the evolution of the real part of the non-diagonal element 
$\rho_{12}={\rm Re}\langle\uparrow\downarrow|\rho|\downarrow\uparrow\rangle$ 
for different system-bath couplings,
with the coupling parametrized by the quantity
$b=\left(\sum_{k=1}^N A^2_k\right)^{1/2}$. 
For every $b$ two curves are shown,
corresponding to $\Gamma_0>\Gamma_{cr}$ (i.e., chaotic bath)
and $\Gamma_0<\Gamma_{cr}$ (regular bath); for the bath here,
$\Gamma_{cr}\sim 0.013$. 
The chaotic bath changes the system's evolution both at
long time (clearly seen on Fig.~\ref{fig1}b), and at short times
(see below). We verified the onset of chaos
by calculating the level spacing statistics $P(s)$ \cite{quchaos},
which agrees with the Wigner-Dyson distribution for chaotic bath
and with Poisson for a regular bath (insets (a),(b) in Fig.~\ref{fig2}).
It is important that other parameters of the bath remain
practically intact: the large-scale structure of the bath's 
spectrum (see Eq.~\ref{bath}) is governed by the local fields 
$h^{x,z}_k$, since $\Gamma_{cr}\ll h_0$. E.g., Fig.~\ref{fig2}
shows that the spectral density of states is practically the same 
for the regular and the chaotic bath. 

\begin{figure}
\includegraphics[width=8cm]{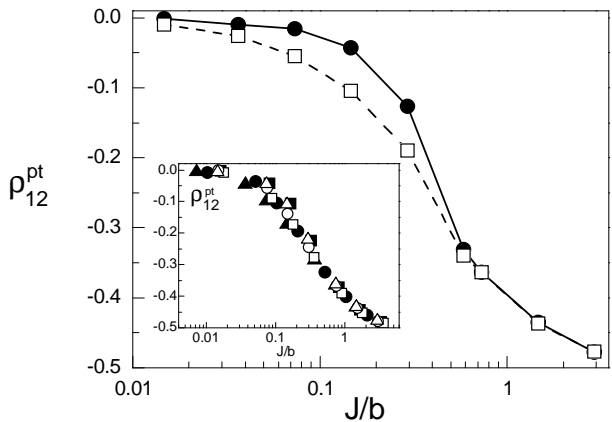}
\caption{\label{fig3} The value $\rho^{pt}_{12}$ vs.\ $J/b$ for 
the chaotic environment $\Gamma_0=0.04, h_0=0.014$ ($\bullet$), and 
for the regular environment $\Gamma_0=0.008, h_0=0,014$ ($\square$). 
The inset shows $\rho^{pt}_{12}$ as a function of $J/b$ for $h_0=1/\sqrt{2}$,
$\Gamma=0.008$ (regular bath), for different sizes $N$ of the environment and
different values $b$ of the coupling between the central system
and the environment: $N=8, b=0.518 (\square)$, $N=10, b=0.562
(\circ)$, $N=12, b=0.683 (\triangle)$, $N=12, b=0.608
(\blacksquare)$, $N=12, b=0.965 (\bullet)$, $N=12, b=1.365
(\blacktriangle)$.}
\end{figure}

Decoherence can be quantified by the system's entropy, concurrence, etc., but 
particular choice does not affect the conclusions. The element 
$\rho_{12}={\rm Re}\langle\uparrow\downarrow|\rho|\downarrow\uparrow\rangle$
is particularly suitable for our model: it has an obvious physical
meaning, and its evolution 
can be understood from the Hamiltonian~(\ref{sys}). 
The coupling $J{\bf S}_1{\bf S}_2$ inside the central system
preserves the initial
singlet correlation between ${\bf S}_1$ and ${\bf S}_2$, thus
steering the value of $\rho_{12}$ towards $-1/2$. The 
system-bath coupling ${\bf S}_1\sum A_k{\bf I}_k$ entangles 
the spin ${\bf S}_1$ with the bath and destroys the correlations
between ${\bf S}_1$ and ${\bf S}_2$, thus leading $\rho_{12}$ towards zero.
Competition between the two tendencies determines the value 
of $\rho_{12}$ at $t\to\infty$. Inset in Fig.~\ref{fig3} shows
that $\rho^{pt}_{12}=\rho_{12}(t\to\infty)$ is determined
by the single ratio $J/b$ (where $b^2=\sum A_k^2$), independently of the size of the bath
$N$ and particular values of $A_k$. But the bath's internal dynamics 
noticeably affects the dependence $\rho^{pt}_{12}(J/b)$, as Fig.~\ref{fig1}b
shows. Fig.~\ref{fig3} presents the results 
of many simulations,
comparing the curves $\rho^{pt}_{12}(J/b)$ for chaotic and
regular environments. The chaotic bath, for the
same value of $J/b$, is more efficient in steering $\rho^{pt}_{12}$
towards zero, i.e.\ in breaking the correlations between 
${\bf S}_1$ and ${\bf S}_2$. 

\begin{figure}
\centering
\includegraphics[width=8cm]{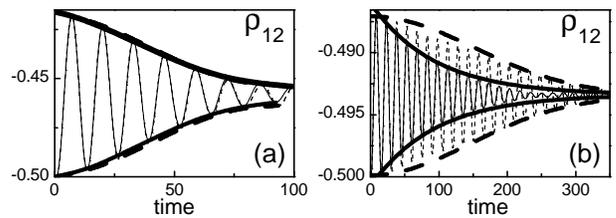}
\caption{\label{fig4} Short-time evolution of the off-diagonal element
$\rho_{12}(t)$ in the case of a chaotic environment
$\Gamma_0=0.04$, $h_0=0.014$ (solid curve) and in the case of a
non-chaotic environment $\Gamma_0=0.008$, $h_0=0,014$ (dashed curve).
The coupling between the two central spins is $J=0.4$. The values
of the interaction between the central system and its environment
are $b=0.2$ (a), and $b=0.0683$ (b). For $b=0.2$ (a), the oscillations
for the chaotic and the regular bath are almost identical, so that
the solid and the dashed curves coincide.}
\end{figure}

\begin{figure}
\includegraphics[width=8cm]{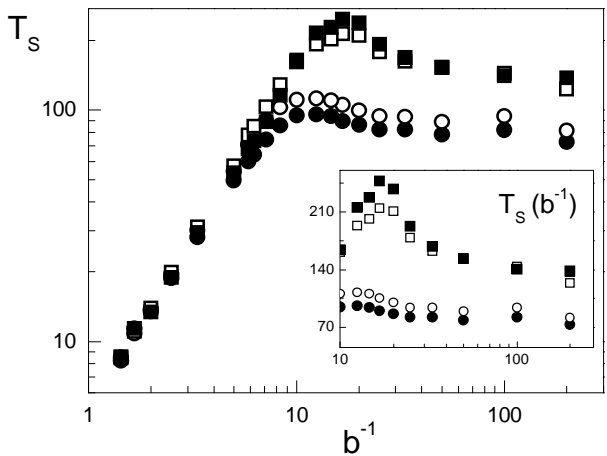}
\caption{\label{fig5} Decoherence time $T_s$ as a function of $1/b$. 
Circles denote the case of a chaotic
environment ($\Gamma_0=0.04$, $h_0=0.014$), squares denote the case of a
non-chaotic environment ($\Gamma_0=0.008$, $h_0=0,014$). To obtain $T_s$
we fitted $\rho^{\rm env}_{12}(t)$ to Gaussian (empty symbols) 
and exponential (solid symbols) forms. 
Inset shows the same data for $1/b> 10$ in a semi-log scale.}
\end{figure}

But the most obvious difference between the chaotic and
the regular baths emerges at short times, $t<100$--300 (Fig.~\ref{fig4}). 
For $J\gg b$ (small system-bath coupling), $\rho_{12}(t)$
oscillates with the frequency $\omega\sim J$, mirroring the quantum 
oscillations of the central system between the singlet and triplet
states. 
For $b$ larger than the spectral width $W$ of the 
bath (here, $W\sim 0.1$, see Fig.~\ref{fig2}), the oscillations of
$\rho_{12}(t)$ are identical for
the chaotic and the regular baths (Fig.~\ref{fig4}a).
The envelope of the oscillations is Gaussian, i.e.\
$\rho^{\rm env}_{12}(t)=\alpha + \beta\exp{\left(-t^2/T_s^2\right)}$ 
where $\alpha$ and $\beta$ are constants, and $T_s$ is the 
decay time. However, when $b$ becomes smaller than $W$, 
prominent differences appear (Fig.~\ref{fig4}b). For the regular bath, the decay 
remains Gaussian, but the chaotic bath leads to the
exponential decay, with the envelope
$\rho^{\rm env}_{12}(t)=\alpha' + \beta'\exp{\left(-t/T_s\right)}$.
The decay time $T_s$ also becomes different, see
Fig.~\ref{fig5}, where $T_s$ is plotted as a function of $1/b$.
The values of $T_s$ were determined from the least-square
fits of $\rho^{\rm env}_{12}(t)$ to both Gaussian and exponential forms;
both forms give similar $T_s$. For $b<W$ (large $1/b$), 
the chaotic-bath decoherence is faster by a factor of 2--2.5
\cite{tsgoesdown}.

It is important to note that the curves $\rho_{12}(t)$ for
the regular bath are insensitive to $\Gamma_0$, but
the drastic difference emerges as soon as $\Gamma_0$ exceeds
$\Gamma_{cr}$, when the bath becomes chaotic.

In our model, the central system can not be considered as a 
Hamiltonian perturbation acting on the bath: ${\bf S}_1$ 
can not be replaced by a fictituous magnetic field 
since the intra-system and the system-bath couplings
are isotropic \cite{emerson}. Also, our bath has no semiclassical limit.
Nonetheless, there is a striking analogy between our 
results and the Loschmidt echo decay in semiclassical systems
\cite{LEgen}. 
Following Ref.~\cite{zurekchaos},
if the central system could be replaced by a
perturbation $\Delta$ of the bath's internal Hamiltonian (\ref{bath}), then
the states $|\uparrow\downarrow\rangle$ and $|\downarrow\uparrow\rangle$ 
of the central system would correspond to
different perturbations $\Delta'$ and $\Delta''$, which
would produce different bath states 
$|\chi'_t\rangle=\exp{(-i[H_B+\Delta']t)}|\chi_0\rangle$ and 
$|\chi''_t\rangle=\exp{(-i[H_B+\Delta'']t)}|\chi_0\rangle$.
The strength of the system-bath interaction $b$ then would correspond 
to the magnitude of $||\Delta||$, and the
matrix element 
$\langle\uparrow\downarrow|\rho|\downarrow\uparrow\rangle$
would correspond to the overlap 
$\langle\chi'_t|\chi''_t\rangle$.
It is known \cite{LEgen} that the quantity 
$F(t)=|\langle\chi'_t|\chi''_t\rangle|^2$ (called Loschmidt echo)
exhibits Gaussian decay when $||\Delta||$ is
larger than the bath's spectral width $W$, and our results for 
$\rho_{12}(t)$ at $b>W$ also show Gaussian decay. 
At $||\Delta||<W$
(in Lyapunov's regime), the Loschmidt echo of the chaotic bath decays 
exponentially with the rate independent of $||\Delta||$, 
while for the regular bath the decay is Gaussian. 
Our simulations give the same picture, with the decay time $T_s$
almost independent of $b$ for chaotic bath ($T_s$ changes by only $\sim 20$\%  
for $0.005<b<0.1$).
So, it is likely that
the $b<W$ regime of decoherence corresponds to the Lyapunov's
regime of the bath, in spite of the fact that our bath
has no semiclassical analog \cite{flambaum}. The study of the
Lyapunov's exponents is an interesting problem for further research.

Summarizing, we compare decoherence of a two-spin system 
by regular and chaotic spin baths. 
We go beyond the standard
one-body semiclassical description, considering environments
of many spins 1/2. We do not replace the system by a perturbation 
acting on the bath, thus going beyond the Loschmidt echo studies.
At $t\to\infty$, the chaotic bath leads to smaller 
values of the system's density matrix element 
$\langle\uparrow\downarrow|\rho|\downarrow\uparrow\rangle$ 
than the regular bath, i.e.\
at long times 
the chaotic bath decoheres the system more efficiently. 
At short times, the chaotic bath leads to faster
decay of quantum oscillations in the system, and changes the
form of the decay from Gaussian to exponential. 
Therefore, the onset of chaos in
the bath drastically changes the decoherence dynamics.
Also, based on the analogy with the Loschmidt echo studies, we
give arguments that the chaotic bath is in the Lyapunov's
regime. 

The authors would like to thank D.~Shepelyansky, B.~Georgeot, 
P.~Jacquod, W.~Zurek, J.-P.~Paz, L.~Viola, F.~Cucchietti, M.~I.~Katsnelson,
and M.~Dzero for helpful discussions.
This work was supported 
by the National Security Agency (NSA) and Advanced
Research and Development Activity (ARDA) under Army Research
Office (ARO) contract DAAD 19-03-1-0132.


\begin{thebibliography}{99}
\bibitem{tsgoesdown} It is interesting that at $0.01<b<0.1$, 
for the regular bath, $T_s$ {\it decreases\/} by a factor of $\sim 2$
with decreasing $b$, i.e.\ the system is decohered faster for
smaller system-bath interaction, while for the chaotic
bath $T_s$ is almost constant (the change is $\sim 13$\%). 
We will discuss this effect elsewhere.
\end{thebibliography}
\end{document}